\documentclass[twocolumn,showpacs,amsmath,amssymb,superscriptaddress,aps,prc]{revtex4-1}
\usepackage[dvips]{graphicx}
\usepackage{epsfig}
\usepackage{amsmath}
\usepackage[dvips]{color}
\newcommand {\beq} {\begin{eqnarray}}
\newcommand {\eeq} {\end{eqnarray}}

\newcommand {\lipb} {\mbox{$^{7}$Li+$^{208}$Pb}}
\newcommand {\lipbi} {\mbox{$^{7}$Li+$^{208}$Pb$\rightarrow ^7$Li$(1/2^-)+^{208}$Pb}}

\begin{document}
\title{Toward a microscopic description of reactions involving exotic nuclei}
\author{P. Descouvemont}
\affiliation{Physique Nucl\'eaire Th\'eorique et Physique Math\'ematique, C.P. 229,
Universit\'e Libre de Bruxelles (ULB), B 1050 Brussels, Belgium}
\affiliation{Instituto de Estudos Avan\c{c}ados da Universidade de S\~{a}o Paulo, 
Caixa Postal 72012, 05508-970, S\~{a}o Paulo, SP, Brazil}
\author{M. S. Hussein}
\affiliation{Instituto de Estudos Avan\c{c}ados da Universidade de S\~{a}o Paulo, 
Caixa Postal 72012, 05508-970, S\~{a}o Paulo, SP, Brazil}
\affiliation{Departamento de F\'isica Matem\'{a}tica,
Instituto de F\'isica da Universidade de S\~{a}o Paulo,
Caixa Postal 66318, 05314-970, S\~{a}o Paulo, SP, Brazil}
\date{\today}
\begin{abstract}
We propose an extension of the Continuum Discretized Coupled Channels (CDCC) method, where the projectile
is described by a microscopic cluster model. This microscopic generalization (MCDCC) only relies on
nucleon-target interactions, and therefore presents an important predictive power. Core excitations
can be included without any further parameter. As an example we investigate the $\lipb$ elastic 
scattering at $E_{lab}=27$ and 35 MeV. The $^7$Li nucleus is known to present an $\alpha+t$ cluster
structure, and is well described by the Resonating Group Method. An excellent agreement is obtained
for the $\lipb$ elastic cross sections, provided that breakup channels are properly included. We also
present an application to inelastic scattering, and discuss future applications of the MCDCC.
\end{abstract}
\maketitle

The study of exotic nuclei is of major interest in current nuclear physics research \cite{BCH93,JRF04,TSK13}. These nuclei present unusual properties, such as a low breakup
threshold and an anomalously large root mean square (rms) radius. Experimentally they are investigated through reactions induced by radioactive beams \cite{BNV13}. The first breakthrough in this field was the discovery of a large radius of the $^{11}$Li isotope \cite{THH85}, and led to the definition and introduction  of "halo" 
nuclei in the nuclear nomenclature. A halo nucleus is considered as a tightly bound core nucleus surrounded by one or two weakly bound nucleons. Thanks to the recent development of experimental facilities, other exotic nuclei, such as $^{6}$He, $^{8}$B and $^{14}$Be, can now be produced with high intensities. In recent years, the effects of low breakup threshold energies have been experimentally studied in various processes involving heavy targets, such as elastic scattering \cite{AML09}, breakup \cite{FCR13}, and fusion \cite{CGD06}. As a general statement, the large rms radius of exotic nuclei has a strong impact on the nucleus-nucleus interaction, as it extends further the range of the nuclear component, and increases couplings to continuum states.

An accurate description of the breakup processes requires high quality reaction models. A scattering model essentially relies on two ingredients: $(i)$ a quantum description of the scattering process; 
$(ii)$ a reliable wave function that faithfully describes the projectile. Optimizing
the description of the projectile, in particular for exotic nuclei, is a crucial issue in reaction models.

At high energies, the Glauber model \cite{Gl59}, using the eikonal approximation \cite{SLY03,CH13} provides an accurate description of various cross sections. Early calculations, based on the adiabatic approximation, were recently extended to include excited or breakup states of the projectile \cite{GBC06, OYI03}. The eikonal approximation provides a significant simplification of the Schr\"odinger equation. This makes it possible to perform two-body and three-body breakup calculations, with a correct treatment of scattering boundary conditions \cite{BCD09,PDB12}.

At low energies (i.e.\ typically around the Coulomb barrier) the eikonal approximation is not valid. 
The low-energy region around the Coulomb barrier is most interesting as quantum barrier tunneling 
effects become relevant. In turn, they induce greater sensitivity of the scattering system to the detailed nature of the couplings.
In this energy regime, the Continuum Discretized Coupled Channel (CDCC) method, originally developed for deuteron-induced reactions \cite{Ra74b}, has proved to be an accurate theoretical tool \cite{KYI86,AIK87}. Since the deuteron can be easily broken up, the theoretical description  of the d+nucleus elastic cross section can be significantly improved  by including couplings to the $p+n$ breakup channels. The CDCC theory is also well adapted to exotic nuclei, owing to their low binding energies.

In standard CDCC calculations, the projectile is described by a two-body structure, where 
the constituents interact through an appropriate potential (fitting, for example, the 
ground-state energy). The internal Hamiltonian is then solved over a basis, and the associated 
eigenstates are used in an expansion of the projectile-target wave functions. Positive-energy 
states are referred to as pseudostates (PS) as they provide an approximation of the 
two-body continuum. In addition to the textbook example d+$^{58}$Ni reaction \cite{MAG09}, other reactions have been recently investigated within this framework (see Ref.\ \cite{YMM12} for a recent
review). The formalism has been extended further to three-body projectiles \cite{MHO04,RAG08} to deal with two-neutron halo nuclei such as $^{6}$He and $^{11}$Li, so-called Borromean nuclei.

These traditional CDCC calculations, however, present shortcomings. The Hamiltonian associated with the system requires optical potentials between the target and the projectile constituents. If optical potentials are in general available for nucleons and $\alpha$ particles, they are often
unknown for heavier nuclei, owing to the lack of data on elastic-scattering cross sections. Then,
approximations must be used, either by scaling optical potentials from neighboring nuclei, or
by evaluating folding potentials. Another limitation comes about from the potential-model description of the projectile. If this approximation is, in most cases, reasonable, it may introduce inaccuracies in the cross section. In particular, core excitations are known to be important in many exotic nuclei, and their effect is absent from most CDCC calculations (see, however, Ref.~\cite{SNT06} where core excitations have been included in the
breakup of $^{11}$Be and $^{17}$C on $^9$Be).

In this Letter, we propose an extension of the CDCC theory, by using a microscopic cluster description of the projectile. In the microscopic CDCC approach (MCDCC), the projectile (with $A_p$ nucleons)
 is described by a many-body Hamiltonian
\beq
H_0=\sum_{i=1}^{A_p} t_i + \sum_{i<j=1}^{A_p} v_{ij},
\label{eq1}
\eeq
where $t_i$ is the kinetic energy operator of nucleon $i$, and $v_{ij}$ is a nucleon-nucleon interaction. 
Hamiltonian such as that of Eq.(\ref{eq1}) is common to all microscopic theories such as the Fermionic Molecular Dynamics (FMD) \cite{NF08}, 
the No-Core Shell Model (NCSM) \cite{NVB00b}, or the Variational Monte-Carlo method (VMC) \cite{PWC04}, to
cite a few. However, a fundamental issue in CDCC calculations is the ability of the model to describe continuum states of the projectile and how they influence the reaction dynamics. 
Recent advances in the NCSM \cite{QN08} and in the Green's Function Monte Carlo method \cite{NPW07} have been successful to describe the continuum, but going beyond nucleon-nucleus systems
remains a complicated task within these approaches.
We use here the cluster approximation, known as the Resonating Group  Method (RGM) \cite{Ho77,DD12}, 
where the treatment of nucleus-nucleus scattering is a direct extension of bound-state
calculations. In the RGM,
an eigenstate of the Hamiltonian (\ref{eq1}) is written as an antisymmetric product of cluster wave functions. This method, and the equivalent Generator Coordinate Method (GCM, \cite{WT77}), have been applied to spectroscopic and scattering properties of many systems (see Ref. \cite{DD12} and references therein). 

\begin{figure}[h]
\begin{center}
\epsfig{file=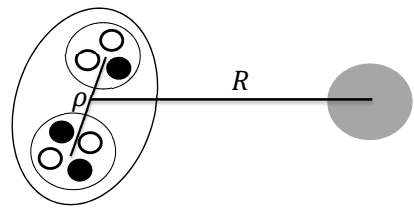,width=0.3\textwidth}
\caption{Schematic picture of the projectile-target system, with a microscopic cluster
structure of the projectile. Coordinates $R$ and $\rho$ are defined in the text.}
\label{fig_config}
\end{center}
\end{figure}

In the present exploratory work, we consider $^{7}$Li as projectile. The RGM-GCM is well known to reproduce most spectroscopic features of this nucleus (as well as of its mirror partner $^{7}$Be), by assuming an $\alpha + t$ structure (or $\alpha + ^{3}$He for the mirror nucleus) \cite{TLT78}. In other words, the RGM $^{7}$Li wave functions associated with $H_0$ are defined as
\beq
\phi^{\ell jm}_k={\mathcal A}\bigl[ [\phi_{\alpha}\otimes \phi_t]^{1/2} \otimes 
Y_{\ell}(\Omega_{\rho})\bigr]^{jm} g^{\ell j}_k(\rho),
\label{eq2}
\eeq
where $\phi_{\alpha}$ and $\phi_t$ are shell model wave functions of the $\alpha$ and $t$ clusters, 
$\ell$ is the orbital momentum, $j$ the total spin, and index $k$ labels the bound and continuum states. In Eq. (\ref{eq2}), $\rho$ is the relative coordinate (see Fig.\ \ref{fig_config}), and  ${\mathcal A}$ is the 7-body antisymmetrization operator which takes into account the Pauli principle among the 7 nucleons of the projectile. The relative wave function $g^{\ell j}_k(\rho)$ are determined from the Schr\"odinger equation associated with $H_0$. 

In general, the RGM equation providing the projectile wave functions (\ref{eq2}) is non local \cite{Ho77}. The GCM is strictly equivalent to the RGM, but is better adapted to numerical calculations, as it makes use of two-cluster Slater determinants. In the GCM, the wave function (\ref{eq2}) is written as
\beq
\phi^{\ell jm}_k=\int f^{\ell j}_k(S)\Phi^{\ell jm}(S) \, dS,
\label{eq3}
\eeq
where $S$ is the generator coordinate, $f^{\ell j}_k(S)$ are the generator functions, 
and $\Phi^{\ell jm}(S)$ are $7\times 7$ projected Slater determinants with four $0s$ orbitals centered at $-3S/7$, and three $0s$ orbitals centered at $4S/7$. Using Slater determinants in the calculation of matrix elements of $H_0$ (and of other operators, such as the electromagnetic ones), is quite systematic, and can be extended to the $p$ and $sd$ shells, even with core excitations
\cite{De96}. Center-of-mass effects are exactly removed when the oscillator parameters of both clusters are identical.

The Hamiltonian of the projectile + target system is defined as
\beq
H=H_0+T_R+\sum_{i=1}^A V_{ti}(\pmb{r}_i-\pmb{R}),
\label{eq4}
\eeq
where $\pmb{R}$ is the projectile-target relative coordinate, $\pmb{r}_i$ the nucleon coordinates
defined from the projectile center of mass (see Fig. \ref{fig_config}) and $V_{ti}$ are the nucleon-target interactions. 
The total wave function is expanded over the GCM projectile basis. For a partial wave with spin $J$ and parity $\pi$, we have
\beq
\Psi^{JM\pi}=\frac{1}{R}
\sum_{cL} \bigl[ \phi^{\ell j}_k\otimes Y_L(\Omega_R)\bigr]^{JM}\, u^{J\pi}_{cL}(R),
\label{eq5}
\eeq
where $L$ is the relative angular momentum, and index $c$ stands for $c = (\ell, j, k)$. 
The summation is truncated at a maximum angular momentum $j_{\rm max}$, and at a maximum PS energy which
limits the number of $k$ values.

The radial wave functions $u^{J\pi}_{cL}(R)$ are obtained from the coupled-channel system 
\beq
&& -\frac{\hbar^2}{2\mu}\biggl[\frac{d^2}{dR^2}-\frac{L(L+1)}{R^2}\biggr]u^{J\pi}_{cL}\nonumber \\
&&+\sum_{c'L'}V^{J\pi}_{cL,c'L'} u^{J\pi}_{c'L'}=(E-E_c)u^{J\pi}_{cL},
\label{eq6}
\eeq
where $E_c$ are the projectile energies, eigenvalues of $H_0$, and where the coupling potentials 
$V^{J\pi}_{cL,c'L'} (R)$ are obtained from the matrix elements
\beq
V_{cc'}(\pmb{R})=\langle \phi^{\ell j}_k \vert \sum_{i=1}^{A_p} V_{ti}(\pmb{r}_i-\pmb{R}) \vert 
\phi^{\ell' j'}_{k'} \rangle,
\label{eq7}
\eeq
and from additional algebraic coefficients. Equation (\ref{eq7}) involves one-body 
matrix elements between Slater determinants $\Phi^{\ell jm}(S)$, which can be computed by using 
the standard formula \cite{Br66}. The system (\ref{eq6}) is then solved 
by using the $R$-matrix method on a Lagrange mesh \cite{DB10,DBD10}. The solutions provide the scattering matrix for all ($J\pi$) values, and consequently various cross sections (elastic and inelastic scattering, breakup, fusion, etc.).

The MCDCC approach presents several advantages: $(1)$ the projectile wave functions are fully antisymmetric, and not limited to bound states; $(2)$ core excitations can be included in a straightforward way; $(3)$ the model only relies on nucleon-target optical potentials. These potentials 
are in general well known, and are independent of the projectile. A strong predictive power of the model is therefore expected.

As mentioned above, our first application of the MCDCC deals with $^{7}$Li elastic scattering on a heavy target, which we take here to be $^{208}$Pb. 
Data are available around the Coulomb barrier ($E_{\rm lab} \approx 30$ MeV) \cite{MGR96}.
As the MCDCC involves heavy numerical calculations, we illustrate the power of the method in a simple case, where $^{7}$Li is described by an $\alpha + t$ cluster structure. The system only involves $0s$ orbitals (with an oscillator parameter $b=1.45$ fm)
and excitations of the $\alpha$ particle can be neglected.

The $^{7}$Li wave functions are defined from a discretization of Eq.~(\ref{eq3}) with 20 values of the generator coordinator $S$, ranging from 0.8 fm to 16 fm in steps of 0.8 fm. The nucleon-nucleon interaction $v_{ij}$ (see Eq.~(\ref{eq1})) is taken as the Minnesota force \cite{TLT78}, complemented with a zero-range spin-orbit term \cite{BP81}. Using the admixture parameter $u = 1.011$, and the spin-orbit amplitude $S_0 =  20.0$ MeV.fm$^5$ reproduces the experimental energies of the $3/2^{-}$ ground state and of the $1/2^{-}$ first excited state simultaneously. 
The $\alpha + t$ wave functions involve partial waves up to $j_{\rm max} = 7/2$ (with both parities).
In addition to the $3/2^-$ and $1/2^-$ bound states of $^7$Li, the $7/2^-$ ($E_{cm}=2.18$ MeV) and $5/2^-$
($E_{cm}=4.13$ MeV)
resonances are also well known cluster states. Continuum states up to 20 MeV are included in the basis. Various tests have been performed to check the stability of the calculated cross sections against the cut-off energy. At the scale of the figures, increasing this energy does not bring any change in
the cross sections.

The present microscopic cluster model is very similar to those used in the past to describe the spectroscopy of $^{7}$Li, the $\alpha+t$ elastic phase shifts, and the $^{3}$H$(\alpha,\gamma)^{7}$Li cross section \cite{Ka86}.
In particular, the quality of the $^7$Li wave functions can be assessed by electromagnetic transition probabilities and by the quadrupole moment of the ground state. For the $B(E1, 3/2^{-} \rightarrow 1/2^{-})$ value, the GCM gives $7.5 \,e^2$.fm$^4$, in good agreement with experiment  $8.3 \pm 0.5 \,e^2.$fm$^4$. The theoretical and experimental values of the ground-state quadrupole moment are $-37.0\,
e$.mb and $-40.6 \pm 0.8\,e$.mb, respectively.

The $^{7}$Li wave functions (including the pseudostates) are then used to determine the 
coupling potentials (\ref{eq7}). The neutron-$^{208}$Pb optical potential (at the neutron energy
of $E_n = E_{\rm lab}/7$) is taken  from
Ref.~\cite{RFW91}, by neglecting the spin-orbit potential. 
The proton-$^{208}$Pb cross section at $E_p = E_{\rm lab}/7$ is virtually identical to the Rutherford cross section \cite{KD03}, and the corresponding interaction only involves the Coulomb potential.

The coupled-channel equations (\ref{eq6}) are then solved with the $R$-matrix method, as alluded to above. For high partial waves, the number of $(cL)$ values can be large (typically up to 150).  Owing to the 
large rms radius of the pseudostates and to the long-range nature of the dipole Coulomb potentials, large channel radii must be used. 
In these conditions the
accuracy of the numerical method is a crucial issue.
Many numerical tests have been performed to check that the cross sections, at the scale of the figures, are not affected by the choice of the channel radius and of the number of basis functions. 
Typical values are 30 fm and 120, respectively.

\begin{figure}[h]
\begin{center}
\epsfig{file=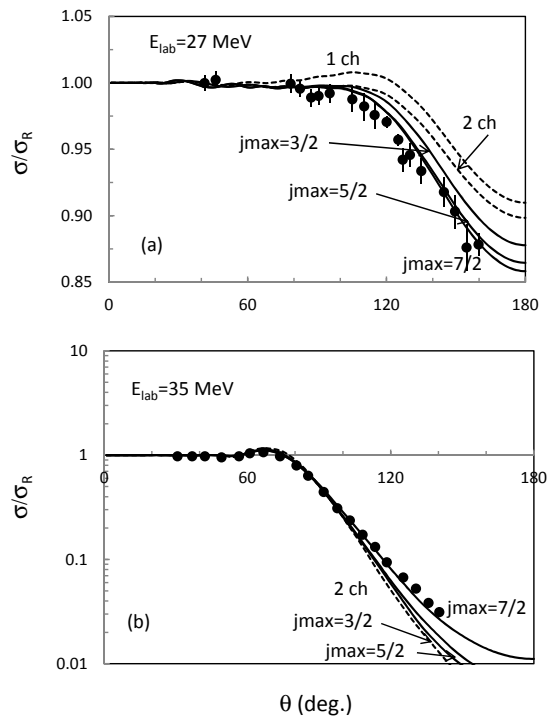,width=0.4\textwidth}
\caption{$\lipb$ elastic cross sections, normalized to the Rutherford cross section,
 at $E_{\rm lab}=27$ MeV (a) and 35 MeV (b). Dotted lines represent the
calculations without breakup channels (at 35 MeV the curves with one and two channels are almost
superimposed), and the solid lines are the full calculations with
increasing $\alpha-t$ angular momentum $j_{\rm max}$. Experimental data are from Ref.~\cite{MGR96}.}
\label{fig_elas}
\end{center}
\end{figure}

In Fig. \ref{fig_elas}, we 
present the elastic-scattering cross sections at $E_{\rm lab} = 27$ and 35 MeV. The calculations 
have been performed by increasing the number of $^{7}$Li states. Obviously, the single-channel 
approach (labeled by ``1 ch''), limited to the $^{7}$Li ground state is not able to reproduce the data. 
At $E_{\rm lab} = 27$ MeV, a slight improvement is obtained by including the $1/2^{-}$ excited state
(labeled by ``2 ch''). 
At both energies, however, an excellent agreement can only be achieved by
including all breakup channels up to $j_{\rm max} = 7/2$. 
This value corresponds to an angular momentum $\ell_{\rm max}=3$ for $j_{\rm max} = 7/2^-$, 
and $\ell_{\rm max}=4$ for $j_{\rm max} = 7/2^+$. Partial wave $j=7/2^-$ is important since it
contains a low-energy resonance. Higher values of the angular momentum are expected to be
negligible.
A non-microscopic CDCC calculation \cite{PJR08} 
requires a renormalization of the $\alpha - ^{208}$Pb and of the $t -^{208}$Pb optical potentials 
by an energy-dependent factor, close to 0.6, which is significantly different from unity; the corresponding cross sections
are therefore strongly affected. Our microscopic approach presents a more powerful predictive procedure, as it does
not contain adjustable parameters.

\begin{figure}[h]
\begin{center}
\epsfig{file=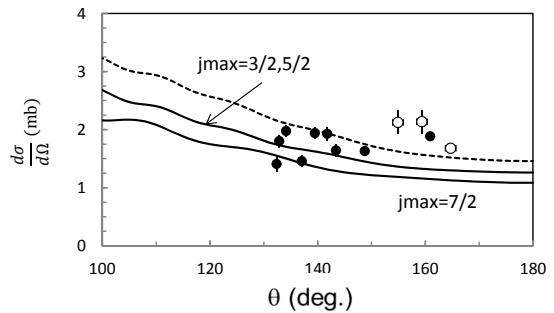,width=0.4\textwidth}
\caption{Inelastic $\lipbi$ cross section at $E_{lab}=27$ MeV. 
The dotted line represents the calculations without breakup channels, i.e. limited to the
ground and first excited states.
The data are taken from Refs.~\cite{MGR98} (black circles) and \cite{PJR08} (open circles). The MCDCC curves for $j_{\rm max}=3/2$ and $j_{\rm max}=5/2$
are superimposed at the scale of the figure.}
\label{fig_inelas}
\end{center}
\end{figure}

Our model can be further tested through the calculation of the inelastic cross section, 
presented in Fig.\ \ref{fig_inelas}. 
At large angles, the nuclear contribution is important.
This cross section is much smaller than the elastic one, and is more sensitive to the details of the wave function. Notwithstanding  that no fitting procedure has been applied, the agreement with the data is fair. Here again, the role of the breakup channels is not negligible. 
In particular, the second excited state $j = 7/2^{-}$, which is a resonant state in the continuum, slightly reduces the cross section.

This exploratory work on the $\lipb$ elastic scattering shows that the MCDCC 
is a powerful tool for the description of low-energy reactions involving weakly bound nuclei, 
where breakup couplings are important. It is expected to be particularly suited  to the scattering 
of exotic nuclei, which present low breakup thresholds, enhancing the effect of 
the continuum. The model is only based on nucleon-target optical potentials, which 
are available over a wide range of masses and scattering energies. Without any 
renormalization factors, we have shown that $\lipb$ elastic and inelastic 
cross sections data can be fairly well reproduced provided that breakup channels 
are properly included.

The present approach opens new perspectives in nucleus-nucleus reaction calculations 
at low energies. We concentrated here on $^{7}$Li, a well known $\alpha - t$ cluster nucleus. 
However, extending Eq. (\ref{eq2}) to include core excitations is quite feasible. In fact, 
several microscopic cluster calculations have been performed with core excitations 
(see, e.g., Ref.~\cite{De97} for $^{11}$Be, and Ref.~\cite{TD10} for $^{17}$C). Calculations 
involving these exotic nuclei are much more involved, but the model itself is identical. 
Besides, the present approach can be easily extended to three-cluster projectiles, such as 
the Borromean two-neutron halo nuclei, $^{6}$He and $^{11}$Li, where RGM wave functions 
are available \cite{DD09,De97b}. Finally, other processes such as breakup and fusion reactions, both of great current interest, can be described by generalizations of the present work.

\section*{Acknowledgments}
This text presents research results of the IAP programme P7/12 initiated by the Belgian-state 
Federal Services for Scientific, Technical and Cultural Affairs. 
P. D. acknowledges the support of F.R.S.-FNRS (Belgium) and of FAPESP (Brazil). M. S. H. acknowledges partial support of the FAPESP, of the CNPq and of the INCT-IQ (Brazilian agencies).


%

\end{document}